\documentclass[aps,pra,reprint,amsmath,amssymb,superscriptaddress,nobibnotes, longbibliography]{revtex4-1}
\usepackage{graphicx,SIunits}
\usepackage{bm}     
\usepackage{hyperref} 
\usepackage{dsfont}    
\usepackage{color}

\usepackage{enumitem}

\usepackage{amsmath}
\usepackage{amsthm}
\usepackage{amssymb}
\usepackage{tikz}

\usetikzlibrary{decorations.pathreplacing}
\usepackage{braket}
\usepackage{dsfont}

\usepackage{color}
\usepackage{lineno}
\usepackage{etoolbox}

\usepackage{multirow}

\begin{document}

\preprint{APS/123-QED}

\title{Universal quantum cloning beyond noncontextual theory}

  \author{Min Namkung}%
      \affiliation{Center for Quantum Technology, Korea Institute of Science and Technology (KIST), Seoul 02792, Korea}

  \author{Hyang-Tag Lim}
      \email{hyangtag.lim@kist.re.kr}
      \affiliation{Center for Quantum Technology, Korea Institute of Science and Technology (KIST), Seoul 02792, Korea}
      \affiliation{Division of Quantum Information, KIST School, Korea University of Science and Technology, Seoul 02792, Korea}

\date{\today}

\begin{abstract}
Quantum theory fundamentally forbids the perfect copying of an arbitrary unknown quantum state, according to a principle known as the no-cloning theorem. Nevertheless, it is possible to construct a deterministic quantum map that produces multiple approximate copies of an unknown quantum state. This task is referred to as universal quantum cloning, further facilitating numerous quantum technologies such as quantum cryptography and quantum communication. In this work, we theoretically verify that the universal quantum cloning cannot be realized within a noncontextual theory, highlighting its intrinsically nonclassical nature. Our verification first {focuses on revealing that} $1\rightarrow2$ cloning scenario {is fully contextual}, and {further covers general examples to observe the contextual behavior of} $N\rightarrow M$ scenario. We believe that our results regarding quantum cloning serve a key role for understanding both quantum foundation and application.
\end{abstract}


\maketitle

\section{Introduction}
Unlike classical information, quantum information cannot be perfectly duplicated, as established by the quantum no-cloning theorem~\cite{w.k.wootters}. This aspect is particularly advantageous for the development of quantum cryptography and quantum communication~\cite{c.h.bennett,n.gisin,c.w.helstrom,g.cariolaro}. First, as {it is impossible to perfectly clone an unknown quantum state}, meaning that eavesdropping is detected by a sender and a receiver, the no-cloning theorem is connected to information-theoretic security of quantum key distribution~\cite{c.h.bennett,n.gisin}. Second, since it is impossible to reduce overlaps among prepared quantum signals by the cloning, one should establish an optimal measurement to identify the {prepared state} with the highest success probability, leading to quantum communication based on state discrimination~\cite{c.w.helstrom,g.cariolaro}. In this context, the no-cloning theorem is of critical importance in understanding fundamental perspectives of quantum information science and technology~\cite{v.scarani}. 

Nevertheless, the quantum theory allows quantum maps that generate approximate copies of an unknown state, including state-dependent~\cite{l.-m.duan,d.bruss} and universal quantum cloning schemes~\cite{d.bruss,v.buzek,f.sciarrino,n.gisin2}. In state-dependent quantum cloning, the input state is assumed to belong to a finite set of possible quantum states, which can be cloned either deterministically with imperfect fidelity~\cite{v.buzek} or probabilistically with perfect fidelity~\cite{l.-m.duan}. While state-dependent quantum cloning is highly useful when the possible input states are restricted to a finite set, universal quantum cloning is of particular interest because it addresses the more general scenario of cloning an arbitrary unknown quantum state. It was theoretically shown that an arbitrary unknown pure state can be transformed into two approximate copies with a maximum fidelity of 5/6~\cite{d.bruss,v.buzek,f.sciarrino}. There is also a general quantum cloner that takes $N$ identical pure states into $M(>N)$ copies with a maximum fidelity of $\frac{M(N+1)+N}{M(N+2)}$~\cite{n.gisin2}. 

Identifying the fundamental quantum resource responsible for quantum cloning schemes can provide deeper insights into how quantum nature enables the aforementioned various quantum cloning schemes. Under this motivation, {contextuality}~\cite{s.kochen,r.w.spekkens} -- a resource for various protocols such as quantum state discrimination~\cite{d.schmid,s.mukherjee,j.shin,k.flatt,m.namkung,k.flatt2}, quantum communication~\cite{s.gupta}, and quantum sensing~\cite{m.lostaglio} -- serves a key role for unveiling the power of the quantum cloning schemes against operational statistical theory~\cite{m.lostaglio2,m.namkung2}, also referred to as noncontextual theory~\cite{r.w.spekkens}. It has been theoretically verified that a noncontextual model does not achieve the fidelity of deterministic state-dependent quantum cloning~\cite{m.lostaglio2}, and the maximum success probability of probabilistic state-dependent quantum cloning~\cite{m.namkung2}. In these two cases, although the performance is outperformed by a quantum model, it is still possible to establish state-dependent cloning tasks within a noncontextual model. Concerning universal quantum cloning~\cite{d.bruss,v.buzek}, it has recently been shown algebraically that there exists a finite set of pure states exhibiting a contextual advantage~\cite{m.doosi}. As the universal quantum cloning covers entire set of (uncountable) pure states, we should verify whether it is possible to reproduce the universal cloning task that clones arbitrary states within the noncontextual theory.

In this work, we theoretically investigate the contextuality underlying universal quantum cloning from an operational perspective. We show that it is impossible to approximately copy arbitrary states in noncontextual theory. This eventually suggests that the universal quantum cloning is more contextual than the state-dependent cloning, in the sense that replicating it is not even allowed within noncontextual framework. This verification starts from the case of producing two approximate copies deterministically~\cite{v.buzek} or stochastically~\cite{f.sciarrino}. {We further study several examples} of producing $M$ copies from $N$ initial states {to observe behavior of the general cloning scenario}~\cite{n.gisin2}. We believe that our theoretical work paves the way for unveiling genuine quantum resources for quantum technologies including quantum cryptography and communication.

\begin{figure}
\centerline{\includegraphics[width=\columnwidth]{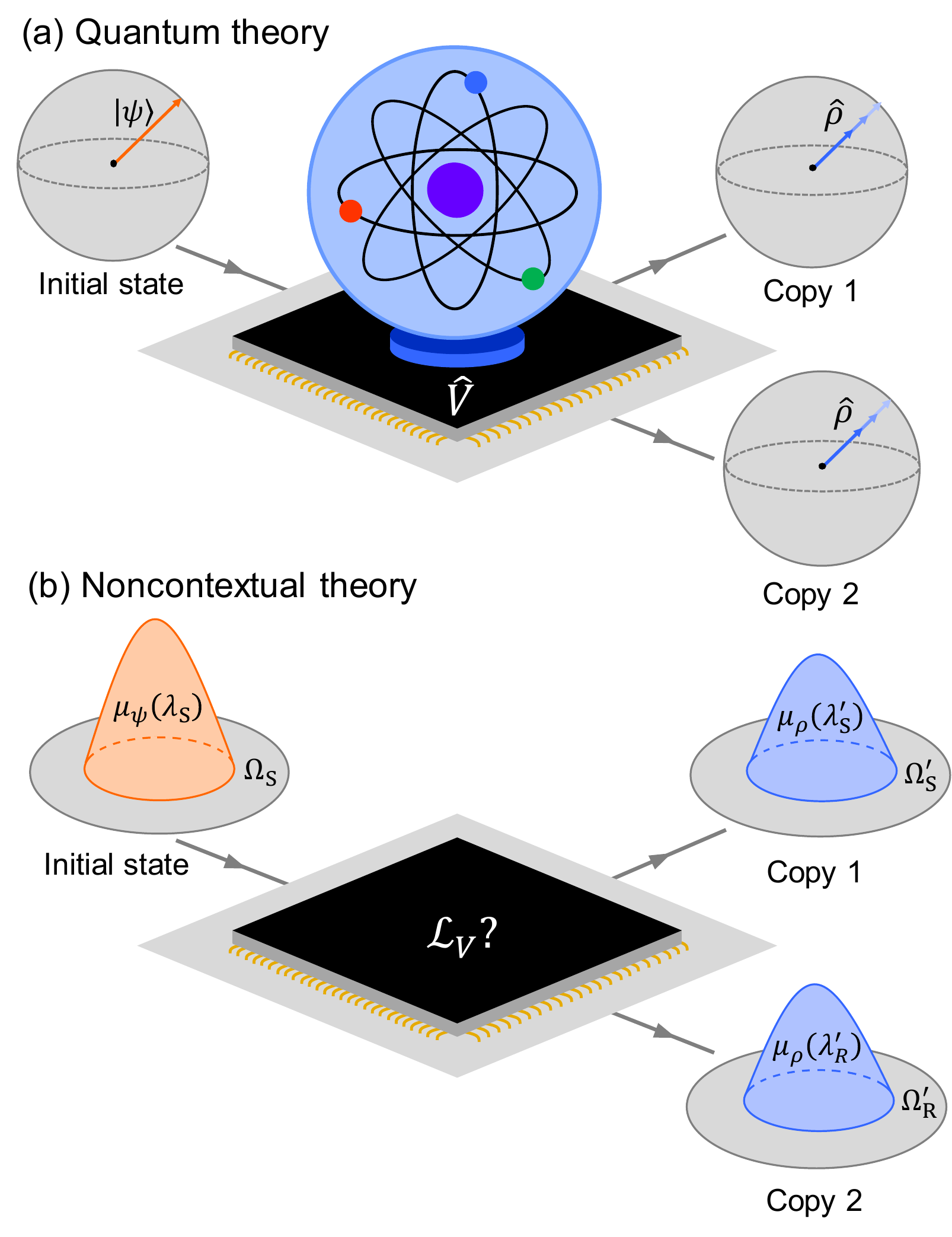}}
\caption{Conceptual figure of a deterministic universal cloning scheme within both quantum and noncontextual theories. (a) Description of the scheme within quantum theory, where approximated copies $\hat{\rho}$ are produced from an unknown pure (qubit) state $|\psi\rangle$ via a unitary operation $\hat{U}$, together with ancilla systems. (b) Description of the scheme within noncontextual theory, where the unknown epistemic state $\mu_\psi(\lambda_{\rm S})$ is copied via a transformation $\mathcal{L}$.}
\centering
\label{fig:1}
\end{figure}

\section{Preliminaries}
We first introduce an operational statistical theory that is called noncontextual theory~\cite{r.w.spekkens}. This theory considers a probability to obtain an outcome $k$ in a prepare-and-measure scenario, denoted by $p(k|\mathcal{P},\mathcal{L},\mathcal{M})$ where $\mathcal{P}$, $\mathcal{L}$, and $\mathcal{M}$ stands for state preparation, transformation, and measurement, respectively. The noncontextual theory has the following operational equivalences regarding state preparation, transformation, and measurement as outlined below:

\begin{enumerate}[label=(\roman*)]
    \item Preparation and measurement equivalence: Two preparations (measurements) $\mathcal{P}$ and $\mathcal{P}'$ ($\mathcal{M}$ and $\mathcal{M}'$)
 are operationally equivalent if they generate identical outcome statistics for all measurements (preparations) and transformations, respectively.
    \item Transformation equivalence: Two transformations $\mathcal{L}$ and $\mathcal{L}'$ are equivalent to each other if $p(k|\mathcal{P},\mathcal{L},\mathcal{M})=p(k|\mathcal{P},\mathcal{L}',\mathcal{M})$ $\forall \mathcal{P},\mathcal{M}$ holds for any outcomes $k$.
\end{enumerate}

From the aforementioned equivalences, an epistemic state $\mu_{\mathcal{P}}(\lambda)$ with respect to preparation $\mathcal{P}$ -- a probability density function of an ontic state $\lambda\in\Omega$ -- is defined, where it satisfies $p(k|\mathcal{P},\mathcal{M})=p(k|\mathcal{P}',\mathcal{M})  \Rightarrow  \mu_\mathcal{P}(\lambda)=\mu_{\mathcal{P}'}(\lambda)$~\cite{k.flatt}. Likewise, response functions $\{\xi_{k|\mathcal{M}}(\lambda)\}$ are defined to establish a measurement, yielding
\begin{eqnarray}
    p(k|\mathcal{P},\mathcal{M})=\int_{\Omega}d\lambda\mu_\mathcal{P}(\lambda)\xi_{k|\mathcal{M}}(\lambda).
\end{eqnarray}

We finally focus on transformation defined as a function $\mathcal{L}(\lambda'|\lambda)$ of ontic states $\lambda\in\Omega$ and $\lambda'$ such that~\cite{r.w.spekkens} 
\begin{eqnarray}
    \widetilde{\mu}_\mathcal{P}(\lambda')=\int_{\Omega}d\lambda\mathcal{L}(\lambda'|\lambda)\mu_\mathcal{P}(\lambda),
\end{eqnarray}
thereby yielding the measurement probability
\begin{eqnarray}
    P(k|\mathcal{P},\mathcal{L},\mathcal{M})=\int_{\Omega'\times\Omega}d\lambda'd\lambda\xi_{k|\mathcal{M}}(\lambda')\mathcal{L}(\lambda'|\lambda)\mu_\mathcal{P}(\lambda).
\end{eqnarray}
Here, $\mathcal{L}$ is assumed to represent a transformation acting on a closed system, meaning that there is no information flow between the system and the environment. This assumption can be theoretically described in terms of the confusability between $\mu_{\mathcal{P}}(\lambda)$ and $\mu_{\mathcal{P}'}(\lambda)$~\cite{d.schmid}:
\begin{eqnarray}
    \Big(\mu_{\mathcal{P}}(\lambda),\mu_{\mathcal{P}'}(\lambda)\Big)_{\Omega}&=&\int_{\mathrm{supp}\left[\mu_\mathcal{P}(\lambda)\right]}d\lambda\mu_{\mathcal{P}'}(\lambda)\nonumber\\
    &=&\int_{\mathrm{supp}\left[\mu_{\mathcal{P}'}(\lambda)\right]}d\lambda\mu_{\mathcal{P}}(\lambda),
\end{eqnarray}
where $\mathrm{supp}\left[\mu_\mathcal{P}(\lambda)\right]$ denotes a support space of $\mu_\mathcal{P}(\lambda)$, defined as a set of ontic states such that $\mu_\mathcal{P}(\lambda)$ is positive-definite. Two epistemic states become more distinguishable (indistinguishable) if the transformation $\mathcal{L}$ increases (decreases) the confusability between them. This physically implies that information regarding the prepared states is flowed inside  (outside) the system, making physical contraction. This leads to that $\mathcal{L}$ should preserve the confusability.

\section{Results}
\subsection{Universal cloning in noncontextual theory}
Before we theoretically verify whether it is possible to establish universal cloning within noncontextual framework as illustrated in Fig.~\ref{fig:1}, we first remind the Bužek-Hillery machine that performs the universal quantum cloning, described by a unitary operator $\hat{V}$ such that~\cite{v.buzek}
\begin{align}\label{uqcm}
    &\hat{V}|\psi\rangle_{\rm S}|0\rangle_{\rm R}|0\rangle_{\rm M}=\sqrt{\frac{2}{3}}|\psi\rangle_{\rm S}|\psi\rangle_{\rm R}|\psi^* \rangle_{\rm M}\\
    & \ \ \ \ \ \ \ \   +\sqrt{\frac{1}{6}}\left(|\psi\rangle_{\rm S}|\psi^\bot\rangle_{\rm R}+|\psi^\bot\rangle_{\rm S}|\psi\rangle_{\rm R}\right)|{\psi^*}^\bot\rangle_{\rm M} \ \  \forall|\psi\rangle,\nonumber
\end{align}
where $|\psi^*\rangle=\alpha^*|0\rangle+\beta^*|1\rangle$ for given $\alpha,\beta\in\mathbb{C}$ and an orthonormal basis $\{|0\rangle,|1\rangle\}$, and $|\psi^\bot\rangle$ ($|{\psi^*}^\bot\rangle$) is a state orthogonal to $|\psi\rangle$ ($|\psi^*\rangle$). The above unitary operation deterministically produces two approximated copies
\begin{eqnarray}\label{app_cop}
    \hat{\rho}_{\rm S}=\hat{\rho}_{\rm R}=\frac{5}{6}|\psi\rangle\langle\psi|+\frac{1}{6}|\psi^\bot\rangle\langle\psi^\bot|.
\end{eqnarray}

Let us subsequently assume that there is a transformation $\mathcal{L}_{V}$ reproducing the universal cloning such that
\begin{align}\label{lv}
    &\int_{\Omega_{\rm S}}\mathcal{L}_V(\lambda_{\rm S}',\lambda_{\rm R}',\lambda_{\rm M}'|\lambda_{\rm S})\mu_\psi(\lambda_{\rm S})=f\mu_{\psi}(\lambda_{\rm S}')\mu_{\psi}(\lambda_{\rm R}')\mu_{\psi^*}(\lambda_{\rm M}') \nonumber\\
    &   +\frac{1-f}{2}\{\mu_{\psi}(\lambda_{\rm S}')\mu_{\psi^\bot}(\lambda_{\rm R}')+\mu_{\psi^\bot}(\lambda_{\rm S}')\mu_{\psi}(\lambda_{\rm R}')\}\mu_{{\psi^*}^\bot}(\lambda_{\rm M}'),
\end{align}
with $f\in[0,1]$ determining the fidelity. Here, $\mu_{\psi}(\lambda')$ and $\mu_{\psi^\bot}(\lambda')$ are epistemic states corresponding to $|\psi\rangle$ and $|\psi^\bot\rangle$, respectively, which are orthogonal to each other in the sense that $\mu_{\psi}(\lambda')\mu_{\psi^\bot}(\lambda')=0$ $\forall\lambda'\in\Omega'$~\cite{d.schmid,k.flatt}. This transformation yields the following two approximated copies
\begin{eqnarray}
    \mu_{\rm S}(\lambda')=\mu_{\rm R}(\lambda')=f\mu_{\psi}(\lambda')+(1-f)\mu_{\psi^\bot}(\lambda').
\end{eqnarray}
Now we propose a main result demonstrating the impossibility of the universal cloning.

\textit{Theorem 1.} There does not exist a transformation $\mathcal{L}_{V}$ that satisfies Eq.~(\ref{lv}) for any initial epistemic states, regardless of the choice of $f$.

\textit{Proof.} Let us assume that there is $\mathcal{L}_{V}$ that fulfills Eq.~(\ref{lv}) for arbitrary epistemic states on $\Omega_{\rm S}$. This means that, for any two epistemic states $\mu_{\psi}(\lambda_{\rm S})$ and $\mu_{\widetilde{\psi}}(\lambda_{\rm S})$ having confusability $c_{\psi,\widetilde{\psi}}$, $\mathcal{L}_{V}$ should satisfy Eq.~(\ref{lv}) for $\mu_{\psi}(\lambda_{\rm S})$ and $\mu_{\widetilde{\psi}}(\lambda_{\rm S})$. We note that $\mathcal{L}_V$ preserves the confusability, leading to the below equality (Detailed derivation is discussed in \textcolor{blue}{Appendix A}):
\begin{eqnarray}\label{10}
    3c_{\psi,\widetilde{\psi}}^2-4c_{\psi,\widetilde{\psi}}+1=0.
\end{eqnarray}
This implies that the confusability between two epistemic states is preserved only when $c_{\psi,\widetilde{\psi}}=\frac{1}{3}$ and $c_{\psi,\widetilde{\psi}}=1$. This spoils the arbitrariness, thereby contradicting the above assumption. \qed

The above theorem suggests that universal cloning is not replicated within a noncontextual framework, unlike several state-dependent cloning schemes are reproduced~\cite{m.lostaglio2,m.namkung2}, albeit sub-optimal.

\begin{figure}
\centerline{\includegraphics[width=\columnwidth]{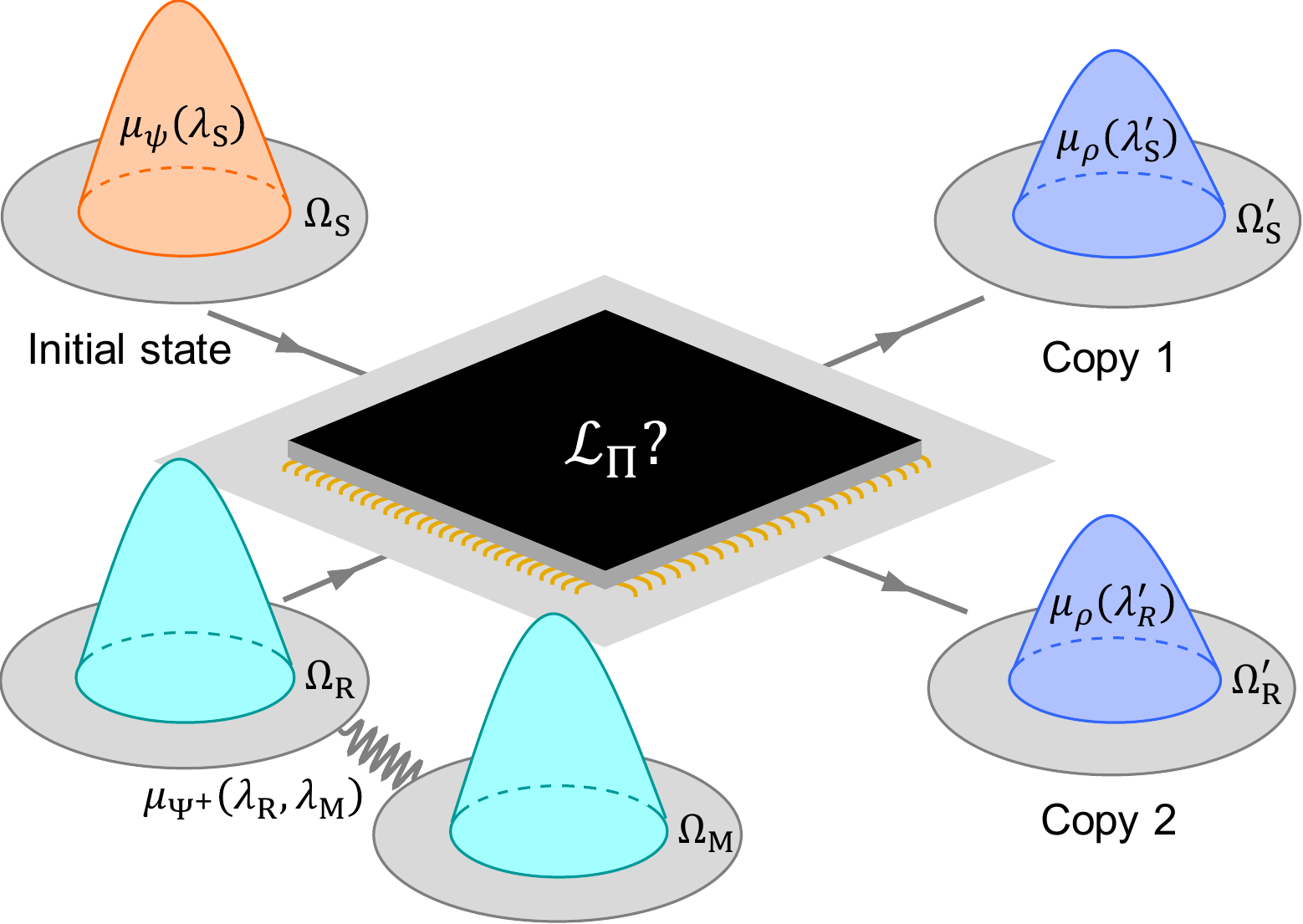}}
\caption{Conceptual figure of postselection-based universal cloning within noncontextual theory. First, an unknown state $\mu_{\psi}(\lambda_{\rm S})$ and $\mu_{\Psi^+}(\lambda_{\rm R},\lambda_{\rm M})$ of Eq.~(\ref{nondet_l}) is prepared, and a nondeterminstic transformation $\mathcal{L}_{\Pi}$ is performed on $\Omega_{\rm S}$ and $\Omega_{\rm R}$, yielding two approximated copies $\mu_{\rho}({\lambda}_{\rm S}')$ and $\mu_{\rho}({\lambda}_{\rm R}')$.}
\centering
\label{fig:2}
\end{figure}

\subsection{Postselection-based universal cloning}
While the Bužek–Hillery quantum cloning machine described by Eq.~(\ref{uqcm}) has been experimentally implemented in several physical platforms~{\cite{f.demartini_,h.k.cummins,z.-b.yang}}, its practical realization can still be challenging in systems where the operation $\hat{V}$ in Eq.~(\ref{uqcm}) is difficult to realize~\cite{a.lamas-linares}. In this case, one can prepare $|\psi\rangle_{\rm S}|\Psi^+\rangle_{\rm R M}$ with a Bell state $|\Psi^\pm\rangle_{\rm RM}=\frac{1}{\sqrt{2}}(|\psi\rangle_{\rm R}|\psi^\bot\rangle_{\rm M}\pm|\psi^\bot\rangle_{\rm R}|\psi\rangle_{\rm M})$, followed by postselection on systems S and R described by $\hat{\Pi}_{\rm SR}=\hat{\mathbb{I}}_{\rm SR}-|\Psi^+\rangle\langle\Psi^+|_{\rm SR}$~\cite{f.sciarrino}. These overall processes yield
\begin{align}\label{theta}
    |\Theta\rangle_{\rm SRM}&=\frac{\left(\hat{\Pi}_{\rm SR}\otimes\hat{\mathbb{I}}_{\rm M}\right)|\psi\rangle_{\rm S}|\Psi^+\rangle_{\rm RM}}{\Vert \left(\hat{\Pi}_{\rm SR}\otimes\hat{\mathbb{I}}_{\rm M}\right)|\psi\rangle_{\rm S}|\Psi^+\rangle_{\rm RM} \Vert}\\
    &=\sqrt{\frac{2}{3}}|\psi\rangle_{\rm S}|\psi\rangle_{\rm R}|\psi^\bot\rangle_{\rm M}+\sqrt{\frac{1}{3}}|\Psi^-\rangle_{\rm SR}|\psi\rangle_{\rm M},\nonumber
\end{align}
with probability $3/4$, whose partial states are equal to Eq.~(\ref{app_cop}). It is noted that performing $\hat{\Pi}_{\rm SR}$ on two states $|\psi\rangle_{\rm S}$ and $\frac{\hat{\mathbb{I}}_{\rm R}}{2}$ with an identity operator $\hat{\mathbb{I}}_{\rm R}$ on a system $\mathrm{R}$ yields the same result.

In this context, one can raise the following question: although deterministic universal cloning is forbidden, is it still possible to reproduce the postselection-based universal cloning in noncontextual theory, as illustrated in Fig.~\ref{fig:2}? We first establish two epistemic states $\mu_{\Psi^+}(\lambda_{\rm R},\lambda_{\rm M})$ and $\mu_{\Phi^+}(\lambda_{\rm R},\lambda_{\rm M})$ analogous to Bell states $|\Psi^+\rangle$ and $|\Phi^+\rangle=\frac{1}{\sqrt{2}}(|\psi\rangle|\psi\rangle+|\psi^\bot\rangle|\psi^\bot\rangle)$~\cite{r.w.spekkens_t}, respectively:
\begin{align}\label{bs}
    \mu_{\Psi^+}(\lambda_{\rm R},\lambda_{\rm M})&=\frac{1}{2}\left\{\mu_\psi(\lambda_{\rm R})\mu_{\psi^\bot}(\lambda_{\rm M})+\mu_{\psi^\bot}(\lambda_{\rm R})\mu_{\psi}(\lambda_{\rm M})\right\},\nonumber\\
    \mu_{\Phi^+}(\lambda_{\rm R},\lambda_{\rm M})&=\frac{1}{2}\left\{\mu_\psi(\lambda_{\rm R})\mu_{\psi}(\lambda_{\rm M})+\mu_{\psi^\bot}(\lambda_{\rm R})\mu_{\psi^\bot}(\lambda_{\rm M})\right\}.\nonumber\\
\end{align}
We remind that epistemic states are probability functions, meaning that it is impossible to establish $\mu_{\Psi^-}(\lambda_{\rm R},\lambda_{\rm M})$ and $\mu_{\Phi^-}(\lambda_{\rm R},\lambda_{\rm M})$ with respect to $|\Psi^-\rangle$ and $|\Phi^-\rangle=\frac{1}{\sqrt{2}}(|\psi\rangle|\psi\rangle-|\psi^\bot\rangle|\psi^\bot\rangle)$. As the two state of Eq.~(\ref{bs}) are orthogonal, they are perfectly distinguished by two response functions 
\begin{align}\label{bm}
    \pi_{\Psi^+}(\lambda_{\rm R},\lambda_{\rm M})&=\pi_\psi(\lambda_{\rm R})\pi_{\psi^\bot}(\lambda_{\rm M})+\pi_{\psi^\bot}(\lambda_{\rm R})\pi_{\psi}(\lambda_{\rm M}),\nonumber\\
    \pi_{\Phi^+}(\lambda_{\rm R},\lambda_{\rm M})&=\pi_\psi(\lambda_{\rm R})\pi_{\psi}(\lambda_{\rm M})+\pi_{\psi^\bot}(\lambda_{\rm R})\pi_{\psi^\bot}(\lambda_{\rm M}),
\end{align}
where $\pi_\psi(\lambda)$ and $\pi_{\psi^\bot}(\lambda)$ are response functions for distinguishing $\mu_\psi(\lambda)$ and $\mu_{\psi^\bot}(\lambda)$. We then construct a nondeterministic transformation 
\begin{eqnarray}\label{nondet_l}
    \mathcal{L}_{\Pi}({\lambda}_{\rm S}',{\lambda}_{\rm R}'|\lambda_{\rm S},\lambda_{\rm R})=\mu_{\Phi^+}({\lambda}_{\rm S}',{\lambda}_{\rm R}')\pi_{\Phi^+}(\lambda_{\rm S},\lambda_{\rm R}),
\end{eqnarray}
analogous to $\hat{\Pi}_{\rm SR}$ in Eq.~(\ref{theta}). Using Eqs.~(\ref{bs})-(\ref{nondet_l}), we finally establish an epistemic state $\mu_\Theta(\widetilde{\lambda}_{\rm S},\widetilde{\lambda}_{\rm R},{\lambda}_{\rm M})$ as
\begin{eqnarray}\label{mu_theta}
    \mu_\Theta({\lambda}_{\rm S}',{\lambda}_{\rm R}',\lambda_{\rm M})\propto\int_{\Omega_{\rm S}\times\Omega_{\rm R}}d\lambda_{\rm S}d\lambda_{\rm R}\mathcal{L}_{\Pi}\mu_\psi\mu_{\Psi^+},
\end{eqnarray}
whose partial states are $\mu_{\mathbb{I}/2}({\lambda}_{\rm S}')$ and $\mu_{\mathbb{I}/2}({\lambda}_{\rm R}')$ corresponding to a maximally mixed state. This means that two approximated copies are not relevant to the original epistemic state, and thus it is meaningless to reproduce postselection-based universal cloning (Detailed discussion is contained in \textcolor{blue}{Appendix B}). 

\textit{Theorem 2.} It is impossible to produce copies other than $\mu_{\mathbb{I}/2}$ using a scheme established in terms of $\mathcal{L}_{\Pi}$ and $\mu_{\Psi^+}$ that constitute $\mu_\Theta$ of Eq.~(\ref{mu_theta}).

\subsection{Generalized universal cloning} 
Beside $1\rightarrow2$ cloning scenario that is described in Eq.~(\ref{uqcm}), we further investigate whether general $N\rightarrow M$ cloning scenarios are reproduced within noncontextual theory. It is known that $N\rightarrow M$ cloning scenarios is performed using the following unitary operation~\cite{n.gisin2}:
\begin{eqnarray}\label{gen}
    \hat{V}_{N,M}|\psi\rangle^{\otimes N}=\sum_{j=0}^{N-M}\alpha_j|(M-j)\psi,j\psi^\bot\rangle\otimes|R_j(\psi)\rangle,
\end{eqnarray}
with symmetric and normalized states $|(M-j)\psi,j\psi^\bot\rangle$ consisting $M-j$ copies of $|\psi\rangle$ and $j$ copies of $|\psi^\bot\rangle$, $|R_j(\psi)\rangle$ denoted by $|R_j(\psi)\rangle=|(M-1-j)\psi^*,j(\psi^*)^\bot\rangle$, and 
\begin{eqnarray}
    \alpha_j=\sqrt{\frac{N+1}{M+1}}\sqrt{\frac{(M-N)!(M-j)!}{(M-N-j)!M!}}.
\end{eqnarray}
We first focus on several examples that lead to the most generalized discussion.

\textit{Example 1.} Let us first assume the case that three approximated copies are produced from one original epistemic state $\mu_\psi(\lambda)$ via deterministic universal cloning. In this case, the confusability-preserving condition of the transformation is given by (the detailed derivation is contained in \textcolor{blue}{Appendix C})
\begin{eqnarray}\label{eg1}
    c_{\psi,\widetilde{\psi}}^4+6c_{\psi,\widetilde{\psi}}^2(1-c_{\psi,\widetilde{\psi}})^2+3(1-c_{\psi,\widetilde{\psi}})^4=1,
\end{eqnarray}
for given two epistemic states $\mu_\psi(\lambda)$ and $\mu_{\widetilde{\psi}}(\lambda)$ having confusability $c_{\psi,\widetilde{\psi}}$. The above equality is satisfied only when $c_{\psi,\widetilde{\psi}}=0.2976807$ and $1$. This means that the considered transformation fails to preserve confusability between any two input states.

\textit{Example 2.} Let us now assume the case that four approximated copies are produced from one original epistemic state $\mu_\psi(\lambda)$. This case corresponds to the confusability-preserving condition
\begin{align}\label{eg2}
    c_{\psi,\widetilde{\psi}}^6+12c_{\psi,\widetilde{\psi}}^4(1-c_{\psi,\widetilde{\psi}})^2+18c_{\psi,\widetilde{\psi}}^2(1-c_{\psi,\widetilde{\psi}})^4+4(1-c_{\psi,\widetilde{\psi}})^6=1.
\end{align}
This has finite solutions $c_{\psi,\widetilde{\psi}}=0.273253$ and $1$ (The detailed derivation is rather lengthy, and thus we refer the reader to follow \textcolor{blue}{Appendix C} for this example), suggesting that it is impossible to construct transformation that preserves confusability between arbitrary two input states.

The above two examples imply that: (i) the confusability-preserving relation of the transformation $\mathcal{L}_{V_{M,N}}$ replicating $\hat{V}_{N,M}$ in Eq.~(\ref{gen}) {is given by $f_{M,N}(c_{\psi,\widetilde{\psi}})=1$ for several $M$ and $N$}, where $f_{M,N}(c_{\psi,\widetilde{\psi}})$ is a polynomial function of $c_{\psi,\widetilde{\psi}}$; and (ii) $f_{M,N}(c_{\psi,\widetilde{\psi}})$ is not always equal to one as observed in Eqs.~(\ref{eg1}) and (\ref{eg2}). These two aspects eventually lead to the {observataion that} $M\rightarrow N$ cloning scenario {with respect to certain $M$ and $N$} is fully contextual.

\section{Conclusion}
We have theoretically demonstrated contextual feature of universal quantum cloning, suggesting that it is impossible to reproduce the cloning schemes within noncontextual framework. Our verification covers not only deterministic universal quantum cloning but also the postselection-based one, and it further extends to {several examples of} $N\rightarrow M$ cloning scenario. Besides, our results are fundamentally different from previous studies on state-dependent and probabilistic quantum cloning, noting that reproducing those two schemes within a noncontextual framework is still possible for those schemes in a sub-optimal sense.

We believe that our results serve as an important step toward identifying the fundamental origin of advantages in quantum cryptography and quantum communication. It was well-known that attacks in quantum cryptography are mainly based on quantum cloning schemes~\cite{v.scarani}. From this perspective, our results imply that cloning-based attacks are impossible without contextuality, suggesting that the security of quantum cryptography is fundamentally related to the contextual nature of quantum physics. It is also known that quantum cloning has application to state estimation~\cite{d.bruss2,j.bae}, meaning that our results may connect the contextuality and the quantum sensing.


\section*{Data availability}
No data were created or analyzed in this study.

\begin{widetext}
\section*{Appendix A. Detailed proof of Theorem 1}
Let us assume that there is a transformation $\mathcal{L}_V$ that performs
\begin{equation}
    \int_{\Omega_{\rm S}}d\lambda_{\rm S}\mathcal{L}_{V}(\lambda_{\rm S}',\lambda_{\rm R}',\lambda_{\rm M}'|\lambda_{\rm S})\mu_{\psi}(\lambda_{\rm S})=\frac{2}{3}\mu_{\psi}(\lambda_{\rm S}')\mu_{\psi}(\lambda_{\rm R}')\mu_{\psi^*}(\lambda_{\rm M}')+\frac{1}{6}\left\{\mu_\psi(\lambda_{\rm S}')\mu_{\psi^\bot}(\lambda_{\rm R}')+\mu_{\psi^\bot}(\lambda_{\rm S}')\mu_{\psi}(\lambda_{\rm R}')\right\}\mu_{{\psi^*}^\bot}(\lambda_{\rm M}'),
\end{equation}
for any initial epistemic states $\mu_{\psi}(\lambda_{\rm S})$, to reproduce optimal universal quantum cloning. Then, for given two epistemic states $\mu_{\psi}(\lambda_{\rm S})$ and $\mu_{\widetilde{\psi}}(\lambda_{\rm S})$ having confusability 
\begin{equation}
    c_{\psi,\widetilde{\psi}}=\big(\mu_{\psi}(\lambda_{\rm S}),\mu_{\widetilde{\psi}}(\lambda_{\rm S})\big)_{\Omega_{\rm S}}=\int_{\mathrm{supp}[\mu_\psi(\lambda_{\rm S})]}d\lambda_{\rm S}\mu_{\widetilde{\psi}}(\lambda_{\rm S})=\int_{\mathrm{supp}[\mu_{\widetilde{\psi}}(\lambda_{\rm S})]}d\lambda_{\rm S}\mu_{\psi}(\lambda_{\rm S}),
\end{equation}
the following two equalities should be satisfied:
\begin{align}\label{trans}
    \int_{\Omega_{\rm S}}d\lambda_{\rm S}\mathcal{L}_{V}(\lambda_{\rm S}',\lambda_{\rm R}',\lambda_{\rm M}'|\lambda_{\rm S})\mu_{\psi}(\lambda_{\rm S})&=\frac{2}{3}\mu_{\psi}(\lambda_{\rm S}')\mu_{\psi}(\lambda_{\rm R}')\mu_{\psi^*}(\lambda_{\rm M}')+\frac{1}{6}\left\{\mu_\psi(\lambda_{\rm S}')\mu_{\psi^\bot}(\lambda_{\rm R}')+\mu_{\psi^\bot}(\lambda_{\rm S}')\mu_{\psi}(\lambda_{\rm R}')\right\}\mu_{(\psi^*)^\bot}(\lambda_{\rm M}'),\nonumber\\
    \int_{\Omega_{\rm S}}d\lambda_{\rm S}\mathcal{L}_{V}(\lambda_{\rm S}',\lambda_{\rm R}',\lambda_{\rm M}'|\lambda_{\rm S})\mu_{\widetilde{\psi}}(\lambda_{\rm S})&=\frac{2}{3}\mu_{\widetilde{\psi}}(\lambda_{\rm S}')\mu_{\widetilde{\psi}}(\lambda_{\rm R}')\mu_{\widetilde{\psi}^*}(\lambda_{\rm M}')+\frac{1}{6}\left\{\mu_{\widetilde{\psi}}(\lambda_{\rm S}')\mu_{\widetilde{\psi}^\bot}(\lambda_{\rm R}')+\mu_{\widetilde{\psi}^\bot}(\lambda_{\rm S}')\mu_{\widetilde{\psi}}(\lambda_{\rm R}')\right\}\mu_{(\widetilde{\psi}^*)^\bot}(\lambda_{\rm M}').
\end{align}
Provided that $\mathcal{L}_V$ describes a closed-system transformation, we obtain the following confusability-preserving equality:
\begin{equation}\label{conf}
    c_{\psi,\widetilde{\psi}}=\left(\int_{\Omega_{\rm S}}d\lambda_{\rm S}\mathcal{L}_{V}(\lambda_{\rm S}',\lambda_{\rm R}',\lambda_{\rm M}'|\lambda_{\rm S})\mu_{\psi}(\lambda_{\rm S}),\int_{\Omega_{\rm S}}d\lambda_{\rm S}\mathcal{L}_{V}(\lambda_{\rm S}',\lambda_{\rm R}',\lambda_{\rm M}'|\lambda_{\rm S})\mu_{\widetilde{\psi}}(\lambda_{\rm S})\right)_{\Omega_{\rm S}'\times\Omega_{\rm R}'\times\Omega_{\rm M}'}.
\end{equation}
Here we note that the support space of $\int_{\Omega_{\rm S}}d\lambda_{\rm S}\mathcal{L}_{V}(\lambda_{\rm S}',\lambda_{\rm R}',\lambda_{\rm M}'|\lambda_{\rm S})\mu_{\psi}(\lambda_{\rm S})$ is represented as
\begin{align}
    &\mathrm{supp}\left[\int_{\mathrm{supp}[\int_{\Omega_{\rm S}}d\lambda_{\rm S}\mathcal{L}_{V}(\lambda_{\rm S}',\lambda_{\rm R}',\lambda_{\rm M}'|\lambda_{\rm S})\mu_{\psi}(\lambda_{\rm S})]}d\lambda_{\rm S}\mathcal{L}_{V}(\lambda_{\rm S}',\lambda_{\rm R}',\lambda_{\rm M}'|\lambda_{\rm S})\mu_{\psi}(\lambda_{\rm S})\right]\nonumber\\
    &=\mathrm{supp}\left[\mu_{\psi}(\lambda_{\rm S}')\mu_{\psi}(\lambda_{\rm R}')\mu_{\psi^*}(\lambda_{\rm M}')+\left\{\mu_\psi(\lambda_{\rm S}')\mu_{\psi^\bot}(\lambda_{\rm R}')+\mu_{\psi^\bot}(\lambda_{\rm S}')\mu_{\psi}(\lambda_{\rm R}')\right\}\mu_{(\psi^*)^\bot}(\lambda_{\rm M}')\right]\nonumber\\
    &=\mathrm{supp}\left[\mu_{\psi}(\lambda_{\rm S}')\mu_{\psi}(\lambda_{\rm R}')\mu_{\psi^*}(\lambda_{\rm M}')\right]\cup\mathrm{supp}\left[\mu_{\psi}(\lambda_{\rm S}')\mu_{\psi^\bot}(\lambda_{\rm R}')\mu_{\psi^*}(\lambda_{\rm M}')\right]\cup \mathrm{supp}\left[\mu_{\psi^\bot}(\lambda_{\rm S}')\mu_{\psi}(\lambda_{\rm R}')\mu_{\psi^*}(\lambda_{\rm M}')\right]\nonumber\\
    &=\Big\{\mathrm{supp}\left[\mu_{\psi}(\lambda_{\rm S}')\right]\times\mathrm{supp}\left[\mu_{\psi}(\lambda_{\rm R}')\right]\times\mathrm{supp}\left[\mu_{\psi^*}(\lambda_{\rm M}')\right]\Big\}\cup\Big\{\mathrm{supp}\left[\mu_{\psi}(\lambda_{\rm S}')\right]\times\mathrm{supp}\left[\mu_{\psi^\bot}(\lambda_{\rm R}')\right]\times\mathrm{supp}\left[\mu_{\psi^*}(\lambda_{\rm M}')\right]\Big\}\nonumber\\
    & \ \ \ \ \ \ \ \ \ \ \ \ \ \ \ \ \ \ \ \ \ \ \ \ \ \ \ \ \ \ \ \ \ \ \ \ \ \ \ \ \ \cup\Big\{\mathrm{supp}\left[\mu_{\psi^\bot}(\lambda_{\rm S}')\right]\times\mathrm{supp}\left[\mu_{\psi}(\lambda_{\rm R}')\right]\times\mathrm{supp}\left[\mu_{\psi^*}(\lambda_{\rm M}')\right]\Big\}.
\end{align}
In the above equation: the first equality holds since omitting the numbers $2/3$ and $1/6$ in Eq.~(\ref{trans}) does not affect the formulation of the support space; the second equality is derived from that support spaces of $\mu_{\psi}(\lambda_{\rm S}')\mu_{\psi}(\lambda_{\rm R}')\mu_{\psi^*}(\lambda_{\rm M}')$, $\mu_{\psi}(\lambda_{\rm S}')\mu_{\psi^\bot}(\lambda_{\rm R}')\mu_{\psi^*}(\lambda_{\rm M}')$, and $\mu_{\psi^\bot}(\lambda_{\rm S}')\mu_{\psi}(\lambda_{\rm R}')\mu_{\psi^*}(\lambda_{\rm M}')$ are orthogonal to each other; and the third equality follows because the three epistemic states in each term are defined on mutually distinct ontic state spaces. Therefore, the right hand side of Eq.~(\ref{conf}) becomes
\begin{align}\label{cal}
    &\left(\int_{\Omega_{\rm S}}d\lambda_{\rm S}\mathcal{L}_{V}(\lambda_{\rm S}',\lambda_{\rm R}',\lambda_{\rm M}'|\lambda_{\rm S})\mu_{\psi}(\lambda_{\rm S}),\int_{\Omega_{\rm S}}d\lambda_{\rm S}\mathcal{L}_{V}(\lambda_{\rm S}',\lambda_{\rm R}',\lambda_{\rm M}'|\lambda_{\rm S})\mu_{\widetilde{\psi}}(\lambda_{\rm S})\right)_{\Omega_{\rm S}'\times\Omega_{\rm R}'\times\Omega_{\rm M}'}\\
    &=\frac{2}{3}\int\limits_{\substack{\{\mathrm{supp}\left[\mu_{\psi}(\lambda_{\rm S}')\right]\times\mathrm{supp}\left[\mu_{\psi}(\lambda_{\rm R}')\right]\times\mathrm{supp}\left[\mu_{\psi^*}(\lambda_{\rm M}')\right]\} 
    \cup\{\mathrm{supp}\left[\mu_{\psi}(\lambda_{\rm S}')\right]\times\mathrm{supp}\left[\mu_{\psi^\bot}(\lambda_{\rm R}')\right]\times\mathrm{supp}\left[\mu_{\psi^*}(\lambda_{\rm M}')\right]\} \\ \{\mathrm{supp}\left[\mu_{\psi^\bot}(\lambda_{\rm S}')\right]\times\mathrm{supp}\left[\mu_{\psi}(\lambda_{\rm R}')\right]\times\mathrm{supp}\left[\mu_{\psi^*}(\lambda_{\rm M}')\right]\}}  
    }d\bm{\lambda}\mu_{\widetilde{\psi}}(\lambda_{\rm S}')\mu_{\widetilde{\psi}}(\lambda_{\rm R}')\mu_{\widetilde{\psi}^*}(\lambda_{\rm M}')\nonumber\\
    &+\frac{1}{6}\int\limits_{\substack{\{\mathrm{supp}\left[\mu_{\psi}(\lambda_{\rm S}')\right]\times\mathrm{supp}\left[\mu_{\psi}(\lambda_{\rm R}')\right]\times\mathrm{supp}\left[\mu_{\psi^*}(\lambda_{\rm M}')\right]\} 
    \cup\{\mathrm{supp}\left[\mu_{\psi}(\lambda_{\rm S}')\right]\times\mathrm{supp}\left[\mu_{\psi^\bot}(\lambda_{\rm R}')\right]\times\mathrm{supp}\left[\mu_{\psi^*}(\lambda_{\rm M}')\right]\} \\ \{\mathrm{supp}\left[\mu_{\psi^\bot}(\lambda_{\rm S}')\right]\times\mathrm{supp}\left[\mu_{\psi}(\lambda_{\rm R}')\right]\times\mathrm{supp}\left[\mu_{\psi^*}(\lambda_{\rm M}')\right]\}}  
    }d\bm{\lambda}\mu_{\widetilde{\psi}}(\lambda_{\rm S}')\mu_{\widetilde{\psi}^\bot}(\lambda_{\rm R}')\mu_{(\widetilde{\psi}^*)^\bot}(\lambda_{\rm M}')\nonumber\\
    &+\frac{1}{6}\int\limits_{\substack{\{\mathrm{supp}\left[\mu_{\psi}(\lambda_{\rm S}')\right]\times\mathrm{supp}\left[\mu_{\psi}(\lambda_{\rm R}')\right]\times\mathrm{supp}\left[\mu_{\psi^*}(\lambda_{\rm M}')\right]\} 
    \cup\{\mathrm{supp}\left[\mu_{\psi}(\lambda_{\rm S}')\right]\times\mathrm{supp}\left[\mu_{\psi^\bot}(\lambda_{\rm R}')\right]\times\mathrm{supp}\left[\mu_{\psi^*}(\lambda_{\rm M}')\right]\} \\ \{\mathrm{supp}\left[\mu_{\psi^\bot}(\lambda_{\rm S}')\right]\times\mathrm{supp}\left[\mu_{\psi}(\lambda_{\rm R}')\right]\times\mathrm{supp}\left[\mu_{\psi^*}(\lambda_{\rm M}')\right]\}}  
    }d\bm{\lambda}\mu_{\widetilde{\psi}^\bot}(\lambda_{\rm S}')\mu_{\widetilde{\psi}}(\lambda_{\rm R}')\mu_{(\widetilde{\psi}^*)^\bot}(\lambda_{\rm M}')\nonumber\\
    &=c_{\psi,\widetilde{\psi}}^3+2c_{\psi,\widetilde{\psi}}(1-c_{\psi,\widetilde{\psi}})^2,\nonumber
\end{align}
with $d\bm{\lambda}=d\lambda_{\rm S}d\lambda_{\rm R}d\lambda_{\rm M}$. Here we used the fact that $|\langle\psi^*|\widetilde{\psi}^*\rangle|$ is equal to $|\langle\psi|\widetilde{\psi}\rangle|$, leading to $c_{\psi^*,\widetilde{\psi}^*}=c_{\psi,\widetilde{\psi}}$. Combining Eqs.~(\ref{conf}) and (\ref{cal}), we finally obtain Eq.~(\ref{10}) of Theorem 1 in the main text. Moreover, this verification holds even when the factor $2/3$ is generalized to any $f\in[0,1]$.

\section*{Appendix B. Detailed proof of Theorem 2}
We first begin the proof by introducing the explicit form of $\mu_{\psi}(\lambda_{\rm S})\mu_{\Psi^+}(\lambda_{\rm R},\lambda_{\rm M})$:
\begin{equation}
    \mu_{\psi}(\lambda_{\rm S})\mu_{\Psi^+}(\lambda_{\rm R},\lambda_{\rm M})=\mu_{\psi}(\lambda_{\rm S})\frac{1}{2}\left\{\mu_\psi(\lambda_{\rm R})\mu_{\psi^\bot}(\lambda_{\rm M})+\mu_{\psi^\bot}(\lambda_{\rm R})\mu_{\psi}(\lambda_{\rm M})\right\}=\mu_{\Theta}(\lambda_{\rm S},\lambda_{\rm R},\lambda_{\rm M}).
\end{equation}
As a nondeterministic transformation $\mathcal{L}_{\Pi}(\lambda_{\rm S}',\lambda_{\rm R}'|\lambda_{\rm S},\lambda_{\rm R})$ will act only on $\Omega_{\rm S}$ and $\Omega_{\rm R}$, it is sufficient to consider the partial state of $\mu_{\Theta}(\lambda_{\rm S},\lambda_{\rm R},\lambda_{\rm M})$:
\begin{equation}
    \mu_{\Theta}(\lambda_{\rm S},\lambda_{\rm R})=\int_{\Omega_{\rm M}}d\lambda_{\rm M}\mu_{\Theta}(\lambda_{\rm S},\lambda_{\rm R},\lambda_{\rm M})=\frac{1}{2}\psi_{\psi}(\lambda_{\rm S})\left\{\mu_{\psi}(\lambda_{\rm R})+\mu_{\psi^\bot}(\lambda_{\rm R})\right\}.
\end{equation}
Under the transformation $\mathcal{L}_{\Pi}(\lambda_{\rm S}',\lambda_{\rm R}'|\lambda_{\rm S},\lambda_{\rm R})$, the epistemic state $\mu_{\Theta}(\lambda_{\rm S},\lambda_{\rm R})$ is transformed into the normalized epistemic state
\begin{equation}
    \frac{\int_{\Omega_{\rm S}\times\Omega_{\rm R}}d\lambda_{\rm S}d\lambda_{\rm R}\mathcal{L}_{\Pi}(\lambda_{\rm S}',\lambda_{\rm R}'|\lambda_{\rm S},\lambda_{\rm R})\mu_{\Theta}(\lambda_{\rm S},\lambda_{\rm R})}{\int_{\Omega_{\rm S}'\times\Omega_{\rm R}'\times\Omega_{\rm S}\times\Omega_{\rm R}}d\lambda_{\rm S}d\lambda_{\rm R}\mathcal{L}_{\Pi}(\lambda_{\rm S}',\lambda_{\rm R}'|\lambda_{\rm S},\lambda_{\rm R})\mu_{\Theta}(\lambda_{\rm S},\lambda_{\rm R})}=\mu_{\Phi^+}(\lambda_{\rm S}',\lambda_{\rm R}'),
\end{equation}
leading to the approximated copies $\mu_{\mathbb{I}/2}(\lambda_{\rm S})$ and $\mu_{\mathbb{I}/2}(\lambda_{\rm R})$. This completes the proof of theorem 2.

\section*{Appendix C. Detailed discussion of Example 1}
We discuss example 1 that concerns the following transformation:
\begin{align}\label{det_eg1}
    &\int_{\Omega_{\rm S}}d\lambda_{\rm S}\mathcal{L}_{V_{1,3}}(\lambda_{\rm S}',\lambda_{\rm R1}',\lambda_{\rm R2}',\lambda_{\rm M1}',\lambda_{\rm M2}'|\lambda_{\rm S})\mu_{\psi}(\lambda_{\rm S})=\frac{1}{2}\mu_{\psi}(\lambda_{\rm S}')\mu_{\psi}(\lambda_{\rm R1}')\mu_{\psi}(\lambda_{\rm R2}')\mu_{\psi^*}(\lambda_{\rm M1}')\mu_{\psi^*}(\lambda_{\rm M2}')\\
    &+\frac{1}{18}\left\{\mu_{\psi}(\lambda_{\rm S}')\mu_{\psi}(\lambda_{\rm R1}')\mu_{\psi^\bot}(\lambda_{\rm R2}')+\mu_{\psi}(\lambda_{\rm S}')\mu_{\psi^\bot}(\lambda_{\rm R1}')\mu_{\psi}(\lambda_{\rm R2}')+\mu_{\psi^\bot}(\lambda_{\rm S}')\mu_{\psi}(\lambda_{\rm R1}')\mu_{\psi}(\lambda_{\rm R2}')\right\}\mu_{\psi^*}(\lambda_{\rm M1}')\mu_{(\psi^*)^\bot}(\lambda_{\rm M2}')\nonumber\\
    &+\frac{1}{18}\left\{\mu_{\psi}(\lambda_{\rm S}')\mu_{\psi}(\lambda_{\rm R1}')\mu_{\psi^\bot}(\lambda_{\rm R2}')+\mu_{\psi}(\lambda_{\rm S}')\mu_{\psi^\bot}(\lambda_{\rm R1}')\mu_{\psi}(\lambda_{\rm R2}')+\mu_{\psi^\bot}(\lambda_{\rm S}')\mu_{\psi}(\lambda_{\rm R1}')\mu_{\psi}(\lambda_{\rm R2}')\right\}\mu_{(\psi^*)^\bot}(\lambda_{\rm M1}')\mu_{\psi^*}(\lambda_{\rm M2}')\nonumber\\
    &+\frac{1}{18}\left\{\mu_{\psi}(\lambda_{\rm S}')\mu_{\psi^\bot}(\lambda_{\rm R1}')\mu_{\psi^\bot}(\lambda_{\rm R2}')+\mu_{\psi^\bot}(\lambda_{\rm S}')\mu_{\psi}(\lambda_{\rm R1}')\mu_{\psi^\bot}(\lambda_{\rm R2}')+\mu_{\psi^\bot}(\lambda_{\rm S}')\mu_{\psi^\bot}(\lambda_{\rm R1}')\mu_{\psi}(\lambda_{\rm R2}')\right\}\mu_{(\psi^*)^\bot}(\lambda_{\rm M1}')\mu_{(\psi^*)^\bot}(\lambda_{\rm M2}')\nonumber,
\end{align}
which reproduces the $1\rightarrow3$ cloning scheme~\cite{n.gisin2}:
\begin{align}
    \hat{V}_{1,3}|\psi\rangle_{\rm S}|R\rangle&=\sqrt{\frac{1}{2}}|\psi\rangle_{\rm S}|\psi\rangle_{\rm R1}|\psi\rangle_{\rm R2}|\psi^*\rangle_{\rm M1}|\psi^*\rangle_{\rm M2}\nonumber\\
    &+\sqrt{\frac{1}{18}}\left(|\psi\rangle_{\rm S}|\psi\rangle_{\rm R1}|\psi^\bot\rangle_{\rm R2}+|\psi\rangle_{\rm S}|\psi^\bot\rangle_{\rm R1}|\psi\rangle_{\rm R2}+|\psi^\bot\rangle_{\rm S}|\psi\rangle_{\rm R1}|\psi\rangle_{\rm R2}\right)(|\psi^*\rangle_{\rm M1}|(\psi^*)^\bot\rangle_{\rm M2}+|(\psi^*)^\bot\rangle_{\rm M1}|\psi^*\rangle_{\rm M2})\nonumber\\
    &+\sqrt{\frac{1}{18}}\left(|\psi\rangle_{\rm S}|\psi^\bot\rangle_{\rm R1}|\psi^\bot\rangle_{\rm R2}+|\psi^\bot\rangle_{\rm S}|\psi\rangle_{\rm R1}|\psi^\bot\rangle_{\rm R2}+|\psi^\bot\rangle_{\rm S}|\psi^\bot\rangle_{\rm R1}|\psi\rangle_{\rm R2}\right)|(\psi^*)^\bot\rangle_{\rm M1}|(\psi^*)^\bot\rangle_{\rm M2}.
\end{align}
The confusability-preserving equality regarding $\mathcal{L}_{V_{1,3}}$ in Eq.~(\ref{det_eg1}) is described by
\begin{align}
    &c_{\psi,\widetilde{\psi}}=\left(\int_{\Omega_{\rm S}}d\lambda_{\rm S}\mathcal{L}_{V_{1,3}}(\lambda_{\rm S}',\lambda_{\rm R1}',\lambda_{\rm R2}',\lambda_{\rm M1}',\lambda_{\rm M2}'|\lambda_{\rm S})\mu_{\psi}(\lambda_{\rm S}),\int_{\Omega_{\rm S}}d\lambda_{\rm S}\mathcal{L}_{V_{1,3}}(\lambda_{\rm S}',\lambda_{\rm R1}',\lambda_{\rm R2}',\lambda_{\rm M1}',\lambda_{\rm M2}'|\lambda_{\rm S})\mu_{\widetilde{\psi}}(\lambda_{\rm S})\right)_{\substack{\Omega_{\rm S}'\times\Omega_{\rm R1}'\times\Omega_{\rm R2}'\\ \times\Omega_{\rm M1}'\times\Omega_{\rm M2}'}}\nonumber\\
    &=\frac{1}{2}\int\limits_{\mathcal{S}}d\bm{\lambda}\mu_{\psi}(\lambda_{\rm S}')\mu_{\psi}(\lambda_{\rm R1}')\mu_{\psi}(\lambda_{\rm R2}')\mu_{\psi^*}(\lambda_{\rm M1}')\mu_{\psi^*}(\lambda_{\rm M2}')\nonumber\\
    &+\frac{1}{18}\int\limits_{\mathcal{S}}d\bm{\lambda}\Big[\left\{\mu_{\psi}(\lambda_{\rm S}')\mu_{\psi}(\lambda_{\rm R1}')\mu_{\psi^\bot}(\lambda_{\rm R2}')+\mu_{\psi}(\lambda_{\rm S}')\mu_{\psi^\bot}(\lambda_{\rm R1}')\mu_{\psi}(\lambda_{\rm R2}')+\mu_{\psi^\bot}(\lambda_{\rm S}')\mu_{\psi}(\lambda_{\rm R1}')\mu_{\psi}(\lambda_{\rm R2}')\right\}\mu_{\psi^*}(\lambda_{\rm M1}')\mu_{(\psi^*)^\bot}(\lambda_{\rm M2}')\nonumber\\
    &+\left\{\mu_{\psi}(\lambda_{\rm S}')\mu_{\psi}(\lambda_{\rm R1}')\mu_{\psi^\bot}(\lambda_{\rm R2}')+\mu_{\psi}(\lambda_{\rm S}')\mu_{\psi^\bot}(\lambda_{\rm R1}')\mu_{\psi}(\lambda_{\rm R2}')+\mu_{\psi^\bot}(\lambda_{\rm S}')\mu_{\psi}(\lambda_{\rm R1}')\mu_{\psi}(\lambda_{\rm R2}')\right\}\mu_{(\psi^*)^\bot}(\lambda_{\rm M1}')\mu_{\psi^*}(\lambda_{\rm M2}')\nonumber\\
    &+\left\{\mu_{\psi}(\lambda_{\rm S}')\mu_{\psi^\bot}(\lambda_{\rm R1}')\mu_{\psi^\bot}(\lambda_{\rm R2}')+\mu_{\psi^\bot}(\lambda_{\rm S}')\mu_{\psi}(\lambda_{\rm R1}')\mu_{\psi^\bot}(\lambda_{\rm R2}')+\mu_{\psi^\bot}(\lambda_{\rm S}')\mu_{\psi^\bot}(\lambda_{\rm R1}')\mu_{\psi}(\lambda_{\rm R2}')\right\}\mu_{(\psi^*)^\bot}(\lambda_{\rm M1}')\mu_{(\psi^*)^\bot}(\lambda_{\rm M2}')\Big],
\end{align}
where $d\bm{\lambda}=d\lambda_{\rm S}'d\lambda_{\rm R1}'d\lambda_{\rm R2}'d\lambda_{\rm M1}'d\lambda_{\rm M2}'$ and 
\begin{align}
    \mathcal{S}&=\mathrm{supp}\left[\int_{\Omega_{\rm S}}d\lambda_{\rm S}\mathcal{L}_{V_{1,3}}(\lambda_{\rm S}',\lambda_{\rm R1}',\lambda_{\rm R2}',\lambda_{\rm M1}',\lambda_{\rm M2}'|\lambda_{\rm S})\mu_{\psi}(\lambda_{\rm S})\right]\nonumber\\
    &=\left\{\mathrm{supp}\left[\mu_{\psi}(\lambda_{\rm S}')\right]\times\mathrm{supp}\left[\mu_{\psi}(\lambda_{\rm R1}')\right]\times\mathrm{supp}\left[\mu_{\psi}(\lambda_{\rm R2}')\right]\times\mathrm{supp}\left[\mu_{\psi^*}(\lambda_{\rm M1}')\right]\times\mathrm{supp}\left[\mu_{\psi^*}(\lambda_{\rm M2}')\right]\right\}\nonumber\\
    &\cup\left\{\mathrm{supp}\left[\mu_{\psi}(\lambda_{\rm S}')\right]\times\mathrm{supp}\left[\mu_{\psi}(\lambda_{\rm R1}')\right]\times\mathrm{supp}\left[\mu_{\psi^\bot}(\lambda_{\rm R2}')\right]\times\mathrm{supp}\left[\mu_{\psi^*}(\lambda_{\rm M1}')\right]\times\mathrm{supp}\left[\mu_{(\psi^*)^\bot}(\lambda_{\rm M2}')\right]\right\}\nonumber\\
    &\cup\left\{\mathrm{supp}\left[\mu_{\psi}(\lambda_{\rm S}')\right]\times\mathrm{supp}\left[\mu_{\psi^\bot}(\lambda_{\rm R1}')\right]\times\mathrm{supp}\left[\mu_{\psi}(\lambda_{\rm R2}')\right]\times\mathrm{supp}\left[\mu_{\psi^*}(\lambda_{\rm M1}')\right]\times\mathrm{supp}\left[\mu_{(\psi^*)^\bot}(\lambda_{\rm M2}')\right]\right\}\nonumber\\
    &\cup\left\{\mathrm{supp}\left[\mu_{\psi^\bot}(\lambda_{\rm S}')\right]\times\mathrm{supp}\left[\mu_{\psi}(\lambda_{\rm R1}')\right]\times\mathrm{supp}\left[\mu_{\psi}(\lambda_{\rm R2}')\right]\times\mathrm{supp}\left[\mu_{\psi^*}(\lambda_{\rm M1}')\right]\times\mathrm{supp}\left[\mu_{(\psi^*)^\bot}(\lambda_{\rm M2}')\right]\right\}\nonumber\\
    &\cup\left\{\mathrm{supp}\left[\mu_{\psi}(\lambda_{\rm S}')\right]\times\mathrm{supp}\left[\mu_{\psi}(\lambda_{\rm R1}')\right]\times\mathrm{supp}\left[\mu_{\psi^\bot}(\lambda_{\rm R2}')\right]\times\mathrm{supp}\left[\mu_{(\psi^*)^\bot}(\lambda_{\rm M1}')\right]\times\mathrm{supp}\left[\mu_{\psi^*}(\lambda_{\rm M2}')\right]\right\}\nonumber\\
    &\cup\left\{\mathrm{supp}\left[\mu_{\psi}(\lambda_{\rm S}')\right]\times\mathrm{supp}\left[\mu_{\psi^\bot}(\lambda_{\rm R1}')\right]\times\mathrm{supp}\left[\mu_{\psi}(\lambda_{\rm R2}')\right]\times\mathrm{supp}\left[\mu_{(\psi^*)^\bot}(\lambda_{\rm M1}')\right]\times\mathrm{supp}\left[\mu_{\psi^*}(\lambda_{\rm M2}')\right]\right\}\nonumber\\
    &\cup\left\{\mathrm{supp}\left[\mu_{\psi^\bot}(\lambda_{\rm S}')\right]\times\mathrm{supp}\left[\mu_{\psi}(\lambda_{\rm R1}')\right]\times\mathrm{supp}\left[\mu_{\psi}(\lambda_{\rm R2}')\right]\times\mathrm{supp}\left[\mu_{(\psi^*)^\bot}(\lambda_{\rm M1}')\right]\times\mathrm{supp}\left[\mu_{\psi^*}(\lambda_{\rm M2}')\right]\right\}\nonumber\\
    &\cup\left\{\mathrm{supp}\left[\mu_{\psi}(\lambda_{\rm S}')\right]\times\mathrm{supp}\left[\mu_{\psi^\bot}(\lambda_{\rm R1}')\right]\times\mathrm{supp}\left[\mu_{\psi^\bot}(\lambda_{\rm R2}')\right]\times\mathrm{supp}\left[\mu_{(\psi^*)^\bot}(\lambda_{\rm M1}')\right]\times\mathrm{supp}\left[\mu_{(\psi^*)^\bot}(\lambda_{\rm M2}')\right]\right\}\nonumber\\
    &\cup\left\{\mathrm{supp}\left[\mu_{\psi^\bot}(\lambda_{\rm S}')\right]\times\mathrm{supp}\left[\mu_{\psi}(\lambda_{\rm R1}')\right]\times\mathrm{supp}\left[\mu_{\psi^\bot}(\lambda_{\rm R2}')\right]\times\mathrm{supp}\left[\mu_{(\psi^*)^\bot}(\lambda_{\rm M1}')\right]\times\mathrm{supp}\left[\mu_{(\psi^*)^\bot}(\lambda_{\rm M2}')\right]\right\}\nonumber\\
    &\cup\left\{\mathrm{supp}\left[\mu_{\psi^\bot}(\lambda_{\rm S}')\right]\times\mathrm{supp}\left[\mu_{\psi^\bot}(\lambda_{\rm R1}')\right]\times\mathrm{supp}\left[\mu_{\psi}(\lambda_{\rm R2}')\right]\times\mathrm{supp}\left[\mu_{(\psi^*)^\bot}(\lambda_{\rm M1}')\right]\times\mathrm{supp}\left[\mu_{(\psi^*)^\bot}(\lambda_{\rm M2}')\right]\right\}
\end{align}
Through the lengthy calculation, we obtain that the confusability-preserving equality is simplified as Eq.~(\ref{eg1}). 
\end{widetext}

\end{document}